\documentclass[aps,prd,a4paper,preprintnumbers,floatfix,nofootinbib]{revtex4}

\usepackage{amssymb}
\usepackage{graphicx}
\usepackage{amsmath, amsthm}
\usepackage{psfrag}
\usepackage{subfigure}
\usepackage{color}
\usepackage{mathrsfs}
\usepackage{graphicx}

\def\be{\begin{equation}}
\def\ee{\end{equation}}
\def\bea{\begin{eqnarray}}
\def\eea{\end{eqnarray}}

\begin{document}
\title{Turnaround radius in scalar-tensor gravity with quasilocal mass}

\author{Valerio Faraoni}
\email[]{vfaraoni@ubishops.ca}
\affiliation{Department of Physics and Astronomy, 
Bishop's University, Sherbrooke, Qu\'ebec, Canada J1M~1Z7
}
\author{Andrea Giusti}
\email[]{agiusti@ubishops.ca}
\affiliation{Department of Physics and Astronomy, 
Bishop's University, Sherbrooke, Qu\'ebec, Canada J1M~1Z7
}

\author{Jeremy C\^ot\'e}
\email[]{jcote16@ubishops.ca}
\affiliation{Perimeter Institute for Theoretical Physics, 31 Caroline 
Street North, Waterloo, ON N2L 2Y5, Canada}

\begin{abstract} 

Following an existing procedure in general relativity, the turnaround 
radius of a spherical structure is studied in scalar-tensor gravity using 
a new prescription for the analog of the Hawking-Hayward quasilocal mass 
in this class of theories.

\end{abstract}



\maketitle

\section{Introduction}
\label{sec:1}

The fact that today we live in an accelerated epoch of the history of the 
universe \cite{Perlmutter:1997zf, Perlmutter:1998np} has several 
consequences. Among these is that there exists an upper limit to 
the size of cosmic structures that are bound. Structures with this size, 
or larger, are dispersed by the cosmic acceleration, while smaller 
structures can remain bound \cite{TR1}-\cite{TR7}. If the structure is 
spherical, the radius at which the local gravitational attraction balances 
exactly the cosmological expansion is known as the {\em turnaround radius} 
\cite{TR1}-\cite{TR7}.

Turnaround physics has been proposed as a probe of dark energy scenarios 
\cite{TR1}-\cite{TR7}. It was soon realized that the turnaround radius and  
turnaround physics are  
also potential probes of the theory of gravity \cite{TRA1}-\cite{TRA7}. 
Spherically symmetric situations have received most of the attention in 
the literature thus far but, recently, more realistic non-spherical  
situations have 
been discussed \cite{Barrow, turnandrea1, turnandrea2, KunzTR, 
GreeksJCAP,Greeksnumerical}. In fact, the observational search for the 
turnaround radius suffers from extrapolating the theory developed in  
oversimplified spherically symmetric scenarios to realistic non-spherical 
ones \cite{Lee:2015upn, Lee:2016oyu, Lee:2017ejv, Lee:2016qpt}.

In the presence of spherical symmetry, the turnaround radius is usually 
defined (both in general relativity (GR) and in alternative gravity) using 
radial timelike geodesics and imposing the condition of zero radial 
acceleration at the turnaround radius. This procedure is, unfortunately, 
gauge-dependent. In GR, an alternative definition uses the Hawking-Hayward 
quasilocal mass \cite{Hawking, Hayward}. In the case of spherical symmetry, to 
which we restrict ourselves in this work, the Hawking-Hayward quasilocal 
energy reduces \cite{Haywardspherical} to the better known 
Misner-Sharp-Hernandez mass \cite{MSH}. In general, the Hawking-Hayward 
quasilocal mass of a cosmological structure described by a perturbation of a 
Friedmann-Lema\^itre-Robertson-Walker (FLRW) universe splits, in a 
gauge-invariant way, into a ``local'' and a ``cosmological'' contribution 
\cite{HHconfo, Nbody, Enzo2013} (the same splitting of the Hawking-Hayward 
quasilocal mass has been applied to study Newtonian dust perturbations in the 
matter era \cite{Nbody} and gravitational lensing by the cosmological constant 
or by dark energy \cite{Lambdalensing, new}). When these two contributions are 
equal (in absolute value), the turnaround radius is achieved 
\cite{AngusMarianne}. In a universe dominated by dark energy with equation of 
state $P_{DE}\simeq -\rho_{DE}$, the numerical value of the turnaround radius 
of a spherical structure obtained with this definition is very close to the 
one obtained with radial timelike geodesics \cite{AngusMarianne}.

When trying to generalize the quasilocal definition of turnaround radius to 
scalar-tensor gravity, the first difficulty that one encounters consists of 
generalizing the Hawking-Hayward quasilocal energy to this class of theories. 
A few tentative definitions have been given in the literature for different 
spacetime geometries or for special situations \cite{otherSTquasilocal}, and 
they disagree with each other. However, one recent proposal relies on minimal 
assumptions, in the sense that it is based simply on writing the scalar-tensor 
field equations as effective Einstein equations, with terms containing the 
Brans-Dicke-like field and its derivatives forming an effective 
energy-momentum tensor \cite{mySTquasilocal}. This procedure is very common in 
scalar-tensor gravity, has been very useful in several contexts ({\em e.g.}, 
\cite{useful1, useful2}), and produces an unambiguous prescription for the 
scalar-tensor analog of the Hawking-Hayward quasilocal mass, which we adopt 
here.

We apply this new generalization of the Hawking-Hayward mass in scalar-tensor 
gravity to the problem of the turnaround radius in cosmology. Restricting 
ourselves to spherical symmetry for simplicity, we use the gauge-invariant 
procedure proposed in Ref.~\cite{AngusMarianne} in the context of GR and we 
characterize the turnaround radius in scalar-tensor gravity. We will begin in 
the context of a perturbed FLRW universe in the conformal Newtonian gauge.

To fix the notation, we recall the basic equations of scalar-tensor gravity. 
We follow the conventions of Ref.~\cite{Wald}: the metric signature is $-+++$ 
and we use units in which the speed  of light and Newton's constant are unity. 
The (Jordan frame) scalar-tensor action is 
\begin{eqnarray}
S_{ST}&=&\int d^4 x \sqrt{-g} \Bigg\{ \Bigg[ 
\frac{1}{16\pi} \left(  \phi {\cal R} 
-\frac{\omega(\phi)}{\phi} \, g^{ab} \nabla_a \phi \nabla_b 
\phi \right)
 -V(\phi) \Bigg]  
+{\cal L}_{(m)} \Bigg\} \,, \label{Jframeaction} 
\end{eqnarray} 
where ${\cal R}$ is the Ricci scalar of the metric $g_{ab}$, which has determinant $g$, while $\phi$ is the Brans-Dicke-like scalar field and, approximately, the inverse of the spacetime-dependent effective gravitational coupling $G_{eff}$. $V(\phi)$ is the potential of this scalar field, while ${\cal L}^{(m)}$ is the Lagrangian density of matter. The field equations read
\begin{eqnarray}
&& R_{ab}-\frac{1}{2} \, g_{ab}{\cal R} = 
\frac{8\pi}{\phi} \, T_{ab} +\frac{\omega}{\phi^2} \left( \nabla_a\phi \nabla_b 
\phi - \frac{1}{2} \, g_{ab} \nabla^c \phi \nabla_c \phi 
\right) +\frac{1}{\phi} \left( \nabla_a\nabla_b \phi 
- g_{ab} \Box \phi \right) -\frac{V}{2\phi} \, g_{ab} 
\,,\label{STfieldeqs} \\
&&\nonumber\\
&&\Box \phi =  \frac{1}{2\omega+3} \left( 8\pi T 
-\frac{d\omega}{d\phi} \, \nabla^c\phi \nabla_c \phi +\phi 
\, \frac{dV}{d\phi} -2V \right)\,,
\end{eqnarray}
where $T_{ab}=-\frac{2}{\sqrt{-g}} \frac{ \delta }{\delta g^{ab}} \left( 
\sqrt{-g}\, {\cal L}_{(m)}\right) $ is the matter energy-momentum tensor with 
trace $T \equiv {T^a}_a$.  $f({\cal R})$ theories of gravity \cite{reviews} 
are a subclass of scalar-tensor theories described by the action
\be
S = \int d^4 x \sqrt{-g} \, f({\cal R}) + 
S^{(m)} \,,
\ee
where $f({\cal R})$ is a nonlinear function of the Ricci scalar. The scalar 
degree of freedom is $\phi = f'({\cal R})$ and the scalar field potential is 
given implicitly by 
\be
V(\phi)= \phi {\cal R}(\phi) -f\left( {\cal R}(\phi)  
\right) \Big|_{\phi=f'( {\cal R}) }  
\ee
in these theories. This action turns out to be equivalent to the 
scalar-tensor one \cite{reviews}
\be
S = \int d^4 x \, \frac{ \sqrt{-g}}{16\pi}  \left[ \phi 
{\cal 
R}-V(\phi) \right] +S^{(m)} \,.
\ee
This is a Brans-Dicke action with vanishing Brans-Dicke parameter  $\omega$ 
and a rather contrived potential $V$ for the Brans-Dicke scalar $\phi$.

\section{Turnaround radius with quasilocal mass in scalar-tensor gravity}
\label{sec:2}

As usual in the literature on the turnaround radius (and on structure 
formation), one does not include vector and tensor perturbations in  
the perturbed FLRW metric, which is justified at lower order for the 
non-relativistic velocities of the matter perturbations. The perturbed FLRW 
metric and Brans-Dicke-like scalar field are
\begin{eqnarray}
ds^2 &= & a^2(\eta)\left[ -\left( 1+2\psi \right)d\eta^2 
+\left(1-2\Phi 
\right) \left( dr^2 +r^2d\Omega_{(2)}^2 \right) \right] \label{lineelement}\\
&&\nonumber\\
&=& g_{\mu\nu} dx^{\mu} dx^{\nu} = I_{\mu\nu} dx^{\mu}dx^{\nu} 
+h_{\mu\nu}dx^{\mu} dx^{\nu}  \,,
\end{eqnarray}
\be
\phi(\eta, r) = \phi_0(\eta) +  \delta\phi(r) \,,
\ee
where $ R^2 d\Omega_{(2)}^2= h_{\mu\nu}dx^{\mu} dx^{\nu} =R^2 \left( 
d\theta^2 +\sin^2 
\theta \, d\varphi^2 \right) $ is the line element on the 2-spheres of 
symmetry with areal radius 
\be
R(\eta, r) = a(\eta)r \sqrt{1-2\Phi} \approx a(\eta)r \left( 1-\Phi 
\right)\equiv R_0 \left( 1-\Phi 
\right) 
\ee
to first order, and
\be
{\cal O}\left( \frac{\delta\phi}{\phi_0} \right) = {\cal O}\left( 
\Phi\right) =
{\cal O}\left( \psi \right) \,.
\ee

The tensors $I_{ab}$ and $h_{ab}$ form the metric $g_{ab}$. In particular, 
the non-zero components $I_{\mu\nu}$ are $0,1$ (the 
time and radial components), while the non-zero 
components $h_{\mu\nu}$ are $2,3$ (the angular ones). 
This simply provides a convenient splitting of the metric for later on.

We also assume that the metric and scalar field perturbations do not depend 
on time, 
\be
\partial_{\eta} \Phi = \partial_{\eta} \psi =  \partial_{\eta} 
\delta\phi=0 \,.
\ee
This assumption is fine for structures of size much smaller than the Hubble 
radius $H^{-1}$, but it would not be justified for 
inflationary perturbations that cross outside the horizon in the early 
universe.\footnote{We thank Enea Romano for bringing this point to our 
attention.} We have in mind applications to the turnaround radius of 
cosmic structures in the late universe. The structures of interest are of the 
size of galaxy groups or clusters, therefore, much smaller than the Hubble 
radius and 
they evolve on time scales much smaller than the Hubble time. For these 
structures, it is justifiable to neglect the time dependence in the metric 
potentials (and, accordingly, in the Brans-Dicke scalar perturbation).

Explicitly, we have (again, keep in mind the range of the indices and that we 
are computing this to first order) 
\begin{eqnarray}
\nonumber I_{\mu\nu} &=& \left(  \begin{array}{cccc}
- (1+2\psi)a^2  & 0 & 0 & 0 \\
 & && \\
0 & (1-2\Phi) a^2 & 0 & 0\\
 & && \\
0 & 0 & 0 & 0 \\
 & && \\
0 & 0 & 0 & 0 
\end{array} \right) \\
&& \nonumber \\
&=& \nonumber I_{\mu\nu}^{(0)} +\delta I_{\mu\nu} \\
&&\nonumber\\
&=&  \left(  \begin{array}{cccc}
- a^2  & 0 & 0& 0\\
 & &&\\
0 &  a^2 & 0 & 0 \\
0 & 0 & 0 & 0 \\
&&&\\
0 & 0 & & 0 
\end{array} \right)
+ \left(  \begin{array}{cccc}
- 2\psi a^2  & 0 & 0\\
 & &&\\
0 & -2\Phi a^2 & 0 & 0 \\
0 & 0 & 0 & 0 \\
&&&\\
0 & 0 & 0 & 0 
\end{array} \right) \,,
\end{eqnarray}

\begin{eqnarray}
\nonumber I^{\mu\nu} &=& \left(  \begin{array}{cccc}
0 & 0 & -\frac{(1-2\Psi)}{a^2} & 0 \\
 &&& \\
0 & 0 & 0 & \frac{(1+2\Phi)}{a^2} \\
0 & 0 & 0 & 0 \\
&&&\\
0 & 0 & 0 & 0 
\end{array} \right)= I^{\mu\nu}_{(0)} +\delta I ^{\mu\nu}\\
&&\nonumber\\
&=& 
\left(  \begin{array}{cccc}
  -\frac{1}{a^2}  & 0 & 0 & 0 \\
 &&& \\
 0 &  \frac{1}{a^2}  &0 & 0 \\
0 & 0 & 0 & 0 \\
&&&\\
0 & 0 & 0 & 0 
\end{array} \right)
+ \left(  \begin{array}{cccc}
 \frac{2\psi}{a^2}  & 0 & 0 & 0 \\
&&& \\
  0 & \frac{2\Phi}{a^2} & 0 & 0 \\
0 & 0 & 0 & 0 \\
&&&\\
0 & 0 & 0 & 0 \\
\end{array} \right) \,,
\end{eqnarray}

\begin{eqnarray}
\nonumber h_{\mu\nu} &=& \left(  \begin{array}{cccc}
0 & 0 & 0 & 0 \\
&&&\\
0 & 0 & 0 & 0 \\
&&&\\
0 & 0 &   a^2(1-2\Phi) r^2 & 0 \\
 &&& \\
0 & 0 & 0 & a^2(1-2\Phi) r^2 \sin^2\theta \\
\end{array} \right) \\
&&\nonumber\\
&=&h_{\mu\nu}^{(0)} +\delta h_{\mu\nu}
=(1-2\Phi) h_{\mu\nu}^{(0)} \,, 
\end{eqnarray}

\begin{eqnarray}
h^{\mu\nu} &=& \left(  \begin{array}{cccc}
0 & 0 & 0 & 0 \\
&&&\\
0 & 0 & 0 & 0 \\
&&&\\
0 & 0 & \frac{(1+2\Phi)}{a^2 r^2} & 0 \\
 &&& \\
0 & 0 &  0 & \frac{(1+2\Phi)}{a^2 r^2 \sin^2\theta} \\
\end{array} \right) \nonumber\\
&&\nonumber\\
&= & h^{\mu\nu}_{(0)} +\delta h^{\mu\nu}=(1+2\Phi) 
h^{\mu\nu}_{(0)} \,.
\end{eqnarray}

The quasilocal mass contained in a 2-sphere of areal radius $R$ in 
scalar-tensor theory, in spherical symmetry, is (cf. Eq.~(20) of 
Ref.~\cite{mySTquasilocal}) 
\begin{eqnarray}
M_{ST} &=& \frac{ \phi R^2}{4} \left[ h^{ac} h^{bc} C_{abcd} 
+\frac{8\pi}{\phi} h^{ab}T_{ab}  -\frac{16\pi T }{3\phi}  
+\frac{\omega}{\phi^2} \left( h^{ab}\nabla_a\phi \nabla_b\phi 
-\frac{1}{3} \, g^{ab}\nabla_a\phi \nabla_b\phi \right) \right. \nonumber\\
&&\nonumber\\
&\, & \left. 
+\frac{ h^{ab} \nabla_a \nabla_b \phi}{\phi} +\frac{V}{3\phi} \right] 
\,,\label{STmass}
\end{eqnarray}
where $C_{abcd}$ is the Weyl tensor and $T \equiv g^{ab}T_{ab}$ is the trace of the matter energy-momentum tensor.

Let us compute the various terms in the right hand side of Eq.~(\ref{STmass}) separately. The first term is
\begin{eqnarray}
h^{ac} h^{bc} C_{abcd} &=& 
h^{22}\left( h^{22} C_{2222} + h^{33} C_{2323}\right) 
+h^{33}\left( h^{22} C_{3232} + h^{33} C_{3333}\right) \nonumber\\
&&\nonumber\\
&=& 2h^{22}_{(0)} h^{33}_{(0)}  C_{2323} = \frac{2 C_{2323}}{a^4 r^4 \sin^2\theta} \,,
\end{eqnarray}
where the Weyl tensor is of first order in the perturbations since it vanishes exactly in the FLRW background. Computing the component $C_{2323}$ gives
\be
h^{ac} h^{bc} C_{abcd} = \frac{2}{3ra^2} \left( \psi' +\Phi' 
 -r\psi'' -r\Phi'' \right) \,,
\ee
where a prime denotes differentiation with respect to the comoving radius $r$. 

The background fluid consists of dark energy and a dust, with total energy density and pressure
\be
\rho_{(0)} = \rho_{(0)}^{(DE)}+\rho_{(0)}^{(dust)} \,, \;\;\;\;\;\;\;\;   
P_{(0)} =P_{(0)}^{(DE)}\,.
\ee
Accordingly, the matter energy-momentum tensor is decomposed as
\begin{widetext}
\begin{eqnarray}
T_{ab}&=& \left(P+\rho \right) u_a u_b +P g_{ab}= 
\left( P_{(0)} +\delta P + \rho_{(0)} +\delta \rho \right)
\left( u^{(0)}_a +\delta u_a \right) \left( u^{(0)}_b +\delta u_b \right) + \left( P_{(0)} + \delta P \right) \left( g_{ab}^{(0)} + \delta g_{ab} \right)  
\nonumber\\
&&\nonumber\\
&=&  \left(P_{(0)}  +\rho_{(0)}  \right) u_a^{(0)}  u_b^{(0)}  +P_{(0)}  
g_{ab}^{(0)} + \left( \delta P +\delta\rho \right) u^{(0)}_a u^{(0)}_b  +
\left( \rho_{(0)} + P_{(0)}\right) \left( u^{(0)}_a \delta u_b +  
u^{(0)}_b \delta u_a \right)
\nonumber\\
&&\nonumber\\
&\, & + P_{(0)} \delta g_{ab} + \delta P 
g_{ab}^{(0)} \nonumber\\
&&\nonumber\\
&\equiv & T_{ab}^{(0)}+\delta T_{ab}\,.
\end{eqnarray}  
\end{widetext}
The non-vanishing components of the matter energy-momentum tensor are
\begin{eqnarray}
T^{(0)}_{00} &=& \rho_{(0)} a^2 \,,\\
&&\nonumber\\
T^{(0)}_{11} &=& a^2 P_{(0)}\,,\\
&&\nonumber\\
T^{(0)}_{22} &=& a^2 P_{(0)}r^2 \,,\\
&&\nonumber\\
T^{(0)}_{33} &=& a^2 P_{(0)}r^2 \sin^2\theta  \,,
\end{eqnarray}
and
\begin{eqnarray}
\delta T_{00} &=& -2a^2 \left( \rho_{(0)} +2P_{(0)} \right) \psi+a^2 
\delta \rho \,,\\
&&\nonumber\\
\delta T_{11} &=& a^2 P_{(0)}
\left( \frac{\delta P}{P_{(0)}} -2\Phi \right)  \,,\\
&&\nonumber\\
\delta T_{22} &=& a^2 r^2 P_{(0)}\left( \frac{\delta P}{P_{(0)}} -2\Phi 
\right)  \,, \\
&&\nonumber\\
\delta T_{33} &=& a^2 r^2 \sin^2\theta  \, P_{(0)} 
\left( \frac{\delta P}{P_{(0)}} -2\Phi \right)  \,,
\end{eqnarray}
so we obtain
\be
8\pi h^{ab}T_{ab} = 16\pi P_{(0)} \left( 1+ \frac{\delta P}{P_{(0)}} 
\right) \,.  
\ee
Then the trace of the energy-momentum tensor is
\begin{eqnarray}
-\frac{16\pi T}{3} &=& -\frac{16\pi}{3} \left( I^{ab}_{(0)}+ \delta I^{ab} 
+ h^{ab}_{(0)} + \delta h^{ab} \right)  \left( T_{ab}^{(0)} + \delta T_{ab} \right) \nonumber\\
&&\nonumber\\
&=& -\frac{16\pi}{3} \left( T^{(0)}+\delta T \right)\,,
\end{eqnarray}
where 
\be
T^{(0)}= g^{ab}_{(0)} T_{ab}^{(0)} 
= -\rho_{(0)}+3P_{(0)} 
\ee
and
\begin{eqnarray}
\delta T &=& I^{ab}_{(0)} \delta T_{ab} + \delta I^{ab}  
T_{ab}^{(0)}  + 
h^{ab}_{(0)} \delta T_{ab} + \delta h^{ab}   
T_{ab}^{(0)} \nonumber\\
&&\nonumber\\
&=& 4\left( \rho_{(0)}+P_{(0)} \right)\psi -\delta \rho + 3\delta P + 
2P_{(0)} \Phi \,,
\end{eqnarray}
so that
\begin{eqnarray}
-\frac{16\pi T}{3} &=& -\frac{16\pi}{3} \left[ -\rho_{(0)}+3P_{(0)} 
+4\left( \rho_{(0)}+P_{(0)} \right)\psi \right.\nonumber\\
&&\nonumber\\
&\, & \left.  -\delta \rho + 
3\delta P + 2P_{(0)} \Phi \right] \,.
\end{eqnarray}
The next term for $M_{ST}$ is 
\be
h^{ab} \nabla_a\phi \nabla_b \phi = 
h^{22} \left( \partial_2 \phi \right)^2 
+ h^{33} \left( \partial_3 \phi \right)^2 =0 
\ee
and, to first order in the perturbations (we expand $\omega$ around the background field $\phi_0$),
\begin{eqnarray}
&& \frac{\omega}{\phi} \left( h^{ab} \nabla_a \phi\nabla_b \phi 
-\frac{1}{3}   
\, g^{ab} \nabla_a \phi\nabla_b \phi \right)  = \frac{\left( \omega_0 -2\omega_0 \psi + d\omega/d\phi|_0 
\right)}{3a^2 
\phi_{(0)} } \left( \phi_{, \eta}^{(0)} \right)^2 \,.
\end{eqnarray}

In order to compute $ h^{ab} \nabla_a\nabla_b \phi $, we need the 
Christoffel symbols 
\begin{eqnarray}
\Gamma^0_{22} &=& \frac{a_{,\eta}}{a}  r^2 \left[1 -2 \left( \Phi+\psi 
\right)\right] \,,\\
&&\nonumber\\
\Gamma^1_{22} &=& -r+  r^2\Phi_{,r}  \,,\\
&&\nonumber\\
\Gamma^0_{33} &=& \frac{a_{,\eta}}{a}  r^2\sin^2 \theta  \left(1- 
2\Phi -2\psi 
\right) \,,\\
&&\nonumber\\
\Gamma^1_{33} &=&  r\sin^2 \theta  \left(-1+r\Phi_{,r} 
\right) \,,\\
&&\nonumber
\end{eqnarray}
and the second covariant derivatives
\begin{eqnarray}
\nabla_2\nabla_2 \phi &=& - \frac{a_{,\eta}}{a}  r^2 \, \phi^{(0)}_{,\eta} 
+\frac{2a_{,\eta}}{a}  r^2 \phi^{(0)}_{,\eta} \left( \Phi+\psi 
\right) +r\delta\phi_{,r} \,, \\
\nabla_3\nabla_3 \phi &=& - \frac{a_{,\eta}}{a}  r^2 \, \sin^2\theta \, 
\phi^{(0)}_{,\eta} + \frac{2a_{,\eta}}{a}  r^2 \, \sin^2\theta 
\,\phi^{(0)}_{,\eta} + r^2 \, \sin^2\theta \, \delta \phi_{,r} \,.
\end{eqnarray}
These expressions yield
\begin{eqnarray}
h^{ab} \nabla_a\nabla_b \phi &=&  \frac{4a_{,\eta}}{a^3} \left(\Phi+\psi 
\right) \phi^{(0)}_{,\eta} - \frac{4a_{,\eta}}{a^3} \phi^{(0)}_{,\eta} 
\Phi  -\frac{2a_{,\eta}}{a^3} \, \phi^{(0)}_{,\eta}
 + \frac{2 \delta\phi_{,r}}{a^2 r} \,,
\end{eqnarray}
while
\be
\frac{V}{3}=\frac{ V( \phi_{(0)}+\delta \phi) }{3}= \frac{ V_0 
+V_0'\delta\phi}{3} \,.
\ee

By inserting these expressions in Eq.~(\ref{STmass}), a few terms 
cancel out and one is left with
\begin{widetext}
\begin{eqnarray}
&& M_{ST}=\frac{R^3}{4} \left[ \frac{16\pi }{3} \, \rho_{(0)} 
+\frac{\omega_0}{3a^2 
\phi_{(0)} } \left( \phi^{(0)}_{,\eta} \right)^2 -\frac{2a_{,\eta}}{a^3} 
\, \phi^{(0)}_{,\eta} +\frac{V_0}{3} \right] \nonumber\\
&&\nonumber\\
&& + \frac{R_{(0)}^3}{4} \left\{ \frac{2\phi^{(0)} C_{2323}}{a^4 r^4 
\sin^2\theta} +\frac{16\pi}{3} \left[-4\left( \rho_{(0)}+P_{(0)} \right) 
\psi +\delta \rho -2P_{(0)}\Phi \right] 
+\frac{ (-2\omega_0 \psi +\omega_0' \delta\phi )}{3a^2 \phi_{(0)} } \left( 
\phi_{,\eta}^{(0)} \right)^2 +\frac{4a_{,\eta}}{ a^3} \, 
\phi^{(0)}_{,\eta} \psi \right.\nonumber\\
&&\nonumber\\
&&\left. 
 +\frac{2\delta\phi_{,r}}{a^2 r} +\frac{V_0'}{3} \, 
\delta \phi \right\} \,. \label{questa}
\end{eqnarray}
\end{widetext}
Using the unperturbed Friedmann equation of scalar-tensor gravity in a 
spatially flat FLRW universe
\be
H_{(0)}^2= \frac{8\pi}{3\phi_{(0)}} \, \rho_{(0)} +\frac{\omega}{6} \left( 
\frac{ \dot{\phi}_{(0)} }{ \phi_{(0)} } \right)^2 -H \,    
\frac{ \dot{\phi}_{(0)} }{ \phi_{(0)} }  +\frac{V_0}{6\phi^{(0)} } 
\ee
(where  an overdot denotes differentiation with respect to the comoving 
time $t$ of the background FLRW universe, related to the conformal time 
$\eta$ by $ dt=ad\eta$), the first square bracket on the right hand side 
of Eq.~(\ref{questa}) becomes $  \frac{R^3}{4} \left[ \, ... \, \right] 
= \frac{H_{(0)}^2 R^3 \phi_{(0)} }{2} $, so that 
\begin{widetext}
\begin{eqnarray}
M_{ST} &=& \frac{H_{(0)}^2R^3 \phi_{(0)}  }{2}
 +\frac{R_{(0)}^3}{4} \left\{ \frac{2\phi^{(0)} C_{2323}}{a^4 r^4 
\sin^2\theta} +\frac{16\pi}{3} \left[-4\left( \rho_{(0)}+P_{(0)} \right) 
\psi +\delta \rho -2P_{(0)}\Phi \right] 
+\frac{ (-2\omega_0 \psi +\omega_0' \delta\phi )}{3a^2 \phi_{(0)} } \left( 
\phi_{,\eta}^{(0)} \right)^2  \right.\nonumber\\
&&\nonumber\\
&\, &\left. +\frac{4a_{,\eta}}{ a^3} \, 
\phi^{(0)}_{,\eta} \psi +\frac{2\delta\phi_{,r}}{a^2 r} +\frac{V_0'}{3} \, 
\delta \phi \right\} \,. \label{questa2}
\end{eqnarray}
Now, using $ R^3=R_{(0)}^3 \left(1-3\Phi \right)$ and  $ M_{ST}^{(0)} = 
H_{(0)}^2 R_{(0)}^3 \phi_{(0)}/2 $, one obtains 
\begin{eqnarray}
M_{ST} &=& M_{ST}^{(0)} \left(1-3\Phi \right) 
 +\frac{R_{(0)}^3 \phi_{(0)} }{4} \left\{ \frac{2}{3 a^2}  
\left( \frac{\psi'}{r} +\frac{\Phi'}{r} - \psi''-\Phi'' \right) 
\right.\nonumber\\
&&\nonumber\\
&\, &\left.
 +\frac{16\pi}{3} \left[-4\left( \rho_{(0)}+P_{(0)} \right) 
\psi +\delta \rho -2P_{(0)}\Phi \right] 
+\frac{ (-2\omega_0 \psi +\omega_0' \delta\phi )}{ 3a^2 \phi_{(0)} } 
\left( \phi_{,\eta}^{(0)} \right)^2  \right.\nonumber\\
&&\nonumber\\
&\, &\left.  + \frac{4a_{,\eta}}{ a^3}  \, 
\phi^{(0)}_{,\eta} \psi +\frac{2\delta\phi_{,r} }{a^2 r} +\frac{V_0'}{3} \, 
\delta \phi \right\} \,. 
\end{eqnarray}
\end{widetext}
The structures interesting for turnaround physics that have been studied 
in the literature are galaxy groups and clusters \cite{Lee:2015upn, 
Lee:2016oyu, Lee:2017ejv, Lee:2016qpt, KunzTR, Greeksnumerical}); they 
have size $\sim R$ much smaller than the Hubble radius $H^{-1}$ and, 
since $|\Phi|\ll 1$, we can simplify: $ M^{(0)}_{ST} \left(1-3\Phi 
\right) = \frac{H_{(0)}^2 R^3_{(0)} \phi_{(0)} }{2} \left(1-3\Phi \right) 
\approx \frac{H_{(0)}^2 R^3_{(0)} \phi_{(0)} }{2} =M^{(0)}_{ST} $ and
\be
M_{ST} \simeq M^{(0)}_{ST}  +\delta M_{ST} \,.
\ee
According to the  procedure of Ref.~\cite{AngusMarianne}, the (comoving) 
turnaround radius is obtained by setting
\be
M^{(0)}_{ST}= \left| \delta M_{ST} \right| \label{newdef}
\ee
or
\begin{eqnarray}
H_{(0)}^2 &=& \frac{2}{3 a^2}  
\left(\frac{\psi'}{r} +\frac{\Phi'}{r} - \psi''-\Phi'' \right) \nonumber\\
&&\nonumber\\
&\,& +\frac{16\pi}{3} \left[-4\left( \rho_{(0)}+P_{(0)} \right) 
\psi +\delta \rho -2P_{(0)}\Phi \right] \nonumber\\
&&\nonumber\\
&\, &
+\frac{ (-2\omega_0 \psi +\omega_0' \delta\phi )}{ 3a^2 \phi_{(0)} } 
\left( \phi_{,\eta}^{(0)} \right)^2 \nonumber\\
&&\nonumber\\
&\, & + \frac{4a_{,\eta}}{ a^3}  \, 
\phi^{(0)}_{,\eta} \psi +\frac{2\delta\phi_{,r} }{a^2 r} +\frac{V_0'}{3} \, 
\delta \phi  \,.\label{finalresult} 
\end{eqnarray} 
This is the equation satisfied by the tunaround radius in scalar-tensor 
gravity.

\section{Comparison with previous literature}
\label{sec:3}

As a check of the previous result, consider the special case of GR in 
which FLRW space, sourced by dark energy with equation 
of state $P_{(0)}=w\rho_{(0)} $ and $w \simeq -1$, is perturbed by a 
point-like mass $m$ and 
\be
\psi=\Phi=- \frac{m}{r} \,,  \;\;\;\;\;\;\;\;\;\;  \phi=\mbox{const.}
\ee
Equation~(\ref{finalresult}) then reduces to
\be
H^2_{(0)} = \frac{2m}{ a^2 r^3}  
+\frac{16\pi}{3} \, \rho_{(0)}  |3w+1|\,  \frac{m}{r} \,. 
\ee
The equations expressing the comoving and areal values of the turnaround 
radius of a spherical structure are
\be
H^2_{(0)} r^3 - 2|3w+2| H_{(0)}^2 mr^2 -\frac{2m}{a^2}= 0\,,
\ee
\be
H_{(0)}^2 R^3 - 2|3w+2| H_{(0)}^2 (ma)R^2 -2(ma)= 0\,,
\ee
respectively. 
Since $ma\ll R $, one can neglect the term proportional to $ma R^2$ in 
comparison with $R^3$, obtaining the turnaround radius
\be
R=\left( \frac{2ma}{H_{(0)}^2} \right)^{1/3} \,,
\ee
which reproduces the well known result for this case, 
corresponding to Eq.~(2.7) of Ref.~\cite{AngusMarianne}.

\section{An example in Brans-Dicke gravity}
\label{sec:3bis}

As an example, we apply the formula derived in Sec.~\ref{sec:2} to a 
family of inhomogeneous, time-dependent solutions of Brans-Dicke theory found 
in \cite{CMB} and interpreted in \cite{ourCMB}. These are exact solutions 
describing spherical inhomogeneities embedded in a FLRW universe sourced by a 
perfect fluid,  but we will linearize them since only Newtonian-like 
perturbations of a FLRW background universe are of interest here.

The general family of solutions of Brans-Dicke theory found in \cite{CMB} 
reads
\be
ds^2 =a^2(\eta) \left[ - A^{2\alpha}(r) d\eta^2 +B(r) \left( dr^2 +r^2 
d\Omega_{(2)}^2 \right) \right]  \label{BDlinelement}
\ee
where
\begin{eqnarray}
A(r) &=& \frac{1-m/(2\alpha r)}{1+m/(2\alpha r)}\,,\\
&&\nonumber\\
B(r) &=& \left( 1+\frac{m}{2\alpha r}\right)^4 
A^{\frac{2(\alpha-1)(\alpha+2)}{\alpha}}  \,,\\
&&\nonumber\\
a(t)&=& a_* \left( \frac{t-t_1}{t_0} \right)^{\frac{ 2\omega 
(2-\gamma)+2}{3\omega \gamma (2-\gamma)+4}} \equiv a_* \tau^{\beta} \,, \label{Eq60}\\
&&\nonumber\\
\phi(t,r) &=& \phi_* \tau^{ \frac{ 2(4-3\gamma)}{3\omega \gamma(2-\gamma)+4} } 
A^{ \frac{-2(\alpha^2-1)}{\alpha}} \,,\\
&&\nonumber\\
\rho(t,r) &=& \rho_* \left( \frac{a_*}{a(t)} \right)^{3\gamma} A^{-2\alpha} 
\,,\\
&&\nonumber\\
\alpha &=& \sqrt{ \frac{2(\omega+2)}{2\omega+3}} \,,
\end{eqnarray}
where $m$ is  a mass parameter, $t_0, t_1, a_*, \phi_*$ and $\rho_*$ are 
constants, 
the comoving time $t$ of the 
``background'' is related to the conformal 
time $\eta$ by $dt=ad\eta$ and we introduced $\beta \equiv 
{\frac{ 2\omega 
(2-\gamma)+2}{3\omega \gamma (2-\gamma)+4}}$, $\tau \equiv \frac{t-t_1}{t_0}$. The energy 
density $\rho$ and pressure $P$ of the cosmic fluid obey the equation of state 
$P=\left( \gamma-1\right) \rho$ with $\gamma=$~const., the Brans-Dicke 
parameter $\omega$ is constant, and there is no potential $V$ for the 
Brans-Dicke scalar field $\phi$. The parameter $\alpha $ is real if 
$\omega<-2$ or if $ \omega >-3/2$. 

Since we are interested in a universe with dust in modified gravity, we set 
$\gamma=1$, then $P_{(0)}=0$ and $\rho_{(0)}(t) \simeq a^{-3}$.   The universe 
approaches a Big Rip if the exponent of 
the scale factor $ \frac{2(\omega+1)}{3\omega+4}<0$, which is satisfied in the 
range $ -4/3 <\omega <-1$ which we adopt. Since here the Brans-Dicke scalar is 
massless, this range is not realistic because it violates the 
Cassini bound $|\omega|>50000$. Nevertheless, we use this solution as 
an example because there are very few inomogeneous solutions of modified 
gravity in the literature and here we want to impose the extra constraints 
that the background FLRW universe is accelerated and is sourced by a dust.

We now linearize the exact solution of \cite{CMB} for $m/r \ll 1$, obtaining
\begin{eqnarray}
A(r) &\simeq & 1-\frac{m}{\alpha \, r} \,,\\
&&\nonumber\\
B(r) &\simeq & 1-\frac{2m (\alpha^2-2)}{\alpha^2 \, r} \,,\\
&&\nonumber\\
\phi(t,r) &=& \phi_* \tau^{\frac{2}{3\omega+4} } \left[ 1+ \frac{2m}{r} 
\left( \frac{\alpha^2-1}{\alpha^2} \right) \right] \,.
\end{eqnarray}
By comparing the linearized line elements~(\ref{lineelement}) and 
(\ref{BDlinelement}), 
one obtains
\begin{eqnarray}
\psi(r) &=& -\frac{m}{r} \,, \label{Eq67}\\
&&\nonumber\\
\Phi(r) &=& \frac{(\alpha^2-2)m}{\alpha^2 r} \,, \label{Eq68}
\end{eqnarray} 
while, for short observation times, one has
\begin{eqnarray}
\phi_{(0)} (t) &=& \phi_* \tau^{\frac{2}{3\omega+4}} \,, \;\;\;\;\;\;\;\;
\delta \phi = \phi_{(0)}(t) \, \frac{2m}{r} \left( \frac{\alpha^2-1}{\alpha^2} 
\right)  \simeq \phi_{(0)}(now) \, \frac{2m}{r} \left( 
\frac{\alpha^2-1}{\alpha^2}  \right) \,,\\
&&\nonumber\\
\rho_{(0)}(t) &=& \frac{\rho_*}{\tau^{3\beta}} \,,\;\;\;\;\;\;\;\;
\delta \rho = \rho_{(0)}(t) \, \frac{2m}{r} 
 \simeq \rho_{(0)}(now) \, \frac{2m}{r} \,.
\end{eqnarray} 
Taking advantage of Eq. \eqref{Eq60} one finds
$$
H_{(0)} =\frac{\beta}{t_0 \tau}
$$ 
and that Eqs. \eqref{Eq67} and \eqref{Eq68} suggest
\be
\frac{ \left( \psi'+\Phi'\right)}{r} -\left( \psi''+\Phi''\right) = 
\frac{6m}{\alpha^2 r^3} 
\, .
\ee
Then, Eq.~(\ref{finalresult}) locating the turnaround radius becomes the cubic 
equation
\be
\alpha_0(t) \, r^3 +\alpha_1 (t)\, r^2 +\alpha_2(t)=0 \,,
\ee
where the coefficients are time-dependent, expressing the fact that the 
turnaround radius evolves with time, and are 
\be
\alpha_0 (t) = 
\left(\frac{\beta}{t_0 \, \tau}\right)^2 
\,,
\ee

\be
\alpha_1 (t)= 
- 2 m
\Bigg[ 
\frac{16 \pi}{3} \Big( \frac{2 \rho_\ast}{\tau^{3 \beta}} + \rho_{(0)}(now) \Big)
- \frac{4 \phi_\ast \, (5 \omega + 6)}{3 \, t_0 ^2 \, (3 \omega + 4)^2} \tau^{-\frac{6 (\omega + 1)}{3 \omega + 4}}
\Bigg]
\,,
\ee

\be
\alpha_2 (t)
=
- \frac{4 m}{a_\ast ^2 \, \tau ^{2 \beta} \, \alpha^2}
\Big[ 1 - (\alpha^2 - 1) \, \phi_{(0)}(now) \Big]
 \,.
\ee

\section{Conclusions}
\label{sec:4}

The turnaround radius in scalar-tensor gravity is now derived using the 
new definition~(\ref{newdef}) already applied to GR in 
\cite{AngusMarianne} to obtain a gauge-invariant expression of the 
turnaround radius. This definition differs from another one based on the 
study of radial timelike geodesics. However, in GR, the final values of 
the turnaround radius obtained with these two different definitions do not 
differ by much \cite{AngusMarianne}, and the observational error in the 
determination of the turnaround radius of realistic structures in the sky 
is going to be more significant than this small difference between 
different definitions. However, the theory needs to be put on a sound 
basis before the significant challenges posed by the observational 
determination of the turnaround radius can be addressed. The new 
definition of scalar-tensor mass generalizing the Hawking-Hayward 
construct to scalar-tensor gravity \cite{mySTquasilocal} produces 
Eq.~(\ref{finalresult}) which locates the turnaround radius in this class 
of theories. In general, this equation appears more complicated than the 
corresponding one of GR due to the presence of the extra scalar degree of 
freedom. There will certainly be simplifications to this formula in 
particular scalar-tensor scenarios which include an accelerated universe 
at the present epoch, to which we do not want to commit at the moment in 
order to preserve generality. The application of the general formula 
(\ref{finalresult}) to detailed scalar-tensor models will be 
considered elsewhere.

\begin{acknowledgments}
We are grateful to Antonio Enea Romano for a discussion. This work is 
supported, in part, by the Natural Science and Engineering Research 
Council of Canada (Grant No. 2016-03803 to V.F.), by Bishop’s 
University, and by the Perimeter Institute for Theoretical Physics. 
Research at Perimeter Institute is supported by the Government of Canada 
through Industry Canada and by the Province of Ontario through the 
Ministry of Economic Development and Innovation.
The work of AG has been carried out in the framework of the
activities of the Italian National Group for Mathematical
Physics [Gruppo Nazionale per la Fisica Matematica
(GNFM), Istituto Nazionale di Alta Matematica (INdAM)].
\end{acknowledgments}



\newpage

\begin{thebibliography}{99}

\bibitem{Perlmutter:1997zf} 
  S.~Perlmutter {\it et al.} [Supernova Cosmology Project Collaboration],
  {\em Nature} {\bf 391}, 51 (1998).

\bibitem{Perlmutter:1998np} 
  S.~Perlmutter {\it et al.} [Supernova Cosmology Project Collaboration],
  {\em Astrophys.\ J.}  {\bf 517}, 565 (1999).

\bibitem{TR1} 
  M.T.~Busha, F.C.~Adams, R.H.~Wechsler, and A.E.~Evrard,
  {\em Astrophys.\ J.} {\bf 596}, 713 (2003).

\bibitem{TR2} 
  V.~Pavlidou and T.N.~Tomaras,
  {\em J. Cosmol. Astropart. Phys.} {\bf 1409}, 020 (2014).

\bibitem{TR3} 
  V.~Pavlidou, N.~Tetradis, and T.N.~Tomaras,
  {\em JCAP} {\bf 1405}, 017 (2014).

\bibitem{TR4} 
S.~Bhattacharya and T.N.~Tomaras,
  {\em Eur. Phys. J. C} {\bf 77}, 526 (2017).

\bibitem{TR5} 
  M.~Cataneo and D.~Rapetti,
 {\em  Int.\ J.\ Mod.\ Phys.\ D} {\bf 27}, 1848006 (2018).

\bibitem{TR7}
  Z.~Roupas,
  {\em Universe} {\bf 5}, 12 (2019).

\bibitem{TRA1}
  V.~Faraoni,
  {\em Phys.\ Dark Universe}  {\bf 11}, 11 (2016).

\bibitem{TRA2}
  S.~Bhattacharya, K.F.~Dialektopoulos, and T.N.~Tomaras,
  {\em J. Cosmol. Astropart. Phys.} {\bf 1605}, 036 (2016).

\bibitem{TRA3}
  S.~Bhattacharya, K.F.~Dialektopoulos, A.E.~Romano, C.~Skordis, and T.N.~Tomaras,
  {\em J. Cosmol. Astropart. Phys.} {\bf 1707}, 018 (2017).

\bibitem{TRA4} 
  S.~Nojiri, S.D.~Odintsov, and V.~Faraoni,
  {\em Phys.\ Rev.\ D} {\bf 98}, 024005 (2018).
   
\bibitem{TRA5} 
  S.~Capozziello, K.F.~Dialektopoulos, and O.~Luongo,
  {\em Int.\ J.\ Mod.\ Phys.\ D} {\bf 28}, 1950058 (2018).

\bibitem{TRA6} 
  R.C.C.~Lopes, R.~Voivodic, L.R.~Abramo and L.~Sodr\'{e},
  {\em J. Cosmol. Astropart. Phys.} {\bf 1809}, 010 (2018). 

\bibitem{TRA7} 
  R.~C.~C.~Lopes, R.~Voivodic, L.~R.~Abramo, and L.~Sodr\'e,
 {\em J. Cosmol. Astropart. Phys.} {\bf 1907}, 026 (2019).

\bibitem{Barrow} J.D. Barrow and J. Silk, {\em Astrophys. J.} {\bf 250}, 
432 (1981); J.D. Barrow and P. Saich, {\em Mon. Not. Roy. Astron. 
Soc.} {\bf 262}, 717 (1993); J.D. Barrow and G. G\"{o}tz, {\em Class. 
Quantum Grav.} {\bf 6}, 1253 (1989).

\bibitem{turnandrea1} A. Giusti and V. Faraoni, {\em Phys. Dark
Universe} {\bf 26}, 100353 (2019).

\bibitem{turnandrea2} A. Giusti and V. Faraoni, arXiv:1911.05130.

\bibitem{KunzTR} 
  S.H.~Hansen, F.~Hassani, L.~Lombriser, and M.~Kunz,
{\em JCAP} {\bf 2001}, 048 (2020).

\bibitem{GreeksJCAP} S. Bhattacharya and T.N.~Tomaras, 
arXiv:1911.06228.

\bibitem{Greeksnumerical} G.~Korkidis, V.~Pavlidou, K.~Tassis, 
E.~Ntormousi, T.N.~Tomaras, and K.~Kovlakas,
arXiv:1912.08216.

\bibitem{Lee:2015upn} 
  J.~Lee, S.~Kim, and S.C.~Rey,
  {\em Astrophys.\ J.}  {\bf 815}, 43 (2015).
   
\bibitem{Lee:2016oyu} 
  J.~Lee and G.~Yepes,
  {\em Astrophys.\ J.} {\bf 832}, 185 (2016).

\bibitem{Lee:2017ejv} 
  J.~Lee,
  {\em Astrophys.\ J.} {\bf 856}, 57 (2018).

\bibitem{Lee:2016qpt} 
J.~Lee,
  {\em Astrophys.\ J.}  {\bf 832}, 123 (2016).

\bibitem{Hawking} S. Hawking, {\em J. Math. Phys. (N.Y.)} 
{\bf 9}, 598 (1968).

\bibitem{Hayward} S.A. Hayward, {\em Phys. Rev. D} {\bf 
49}, 831 (1994).

\bibitem{Haywardspherical} S.A. Hayward, {\em Phys. Rev. D} 
{\bf 53}, 1938 (1996).

\bibitem{MSH} C.W. Misner and D.H. Sharp, {\em Phys. Rev.} 
{\bf 136}, B571 (1964); W.C. Hernandez and C.W. Misner, 
{\em Astrophys. J.} {\bf 143}, 452 (1966).

\bibitem{HHconfo} A. Prain, V. Vitagliano, V. Faraoni, and 
M. Lapierre-L\'eonard, {\em Classical Quantum Grav.} {\bf 
33}, 145008 (2016).

\bibitem{Nbody} V. Faraoni, M. Lapierre-L\'eonard, and A. 
Prain, {\em Phys. Rev. D} {\bf 92}, 023511 (2015).

\bibitem{Enzo2013} V. Faraoni and V. Vitagliano, {\em Phys. 
Rev. D} {\bf 89}, 064015 (2014).

\bibitem{Lambdalensing} V. Faraoni and M. Lapierre-L\'eonard, {\em Phys. 
Rev. D} {\bf 95}, 023509 (2017).

\bibitem{new} M. Lapierre-L\'eonard, V. Faraoni, and F. Hammad, {\em Phys. 
Rev. D} {\bf 96}, 083525 (2017).
 
\bibitem{AngusMarianne} V. Faraoni, M. Lapierre-L\'eonard, and A. Prain, 
{\em J. Cosmol. Astropart. Phys.} {\bf 10}, 013 (2015).

\bibitem{otherSTquasilocal} R.-G. Cai, L.M. Cao, Y.P. Hu, and N. Ohta, 
{\em Phys. Rev. D} {\bf 80}, 104016 (2009); R.-G. Cai, L.M. Cao, Y.P. Hu, 
and S.P. Kim, {\em Phys. Rev. D} {\bf 78}, 124012 (2008); H. Zhang, Y. Hu, 
and X.Z. Li, {\em Phys. Rev. D} {\bf 90}, 024062 (2014); S.-F. Wu, B. 
Wang, and G.-H. Yang, {\em Nucl. Phys. B} {\bf 799}, 330 (2008);  G. 
Cognola, O. Gorbunova, L. Sebastiani, and S. Zerbini, {\em Phys. Rev. D} 
{\bf 84}, 023515 (2011); F. Hammad, {\em Classical Quantum Grav.} {\bf 
33}, 235016 (2016); {\em Int. J. Mod. Phys. D} {\bf 25}, 1650081 (2016).

\bibitem{mySTquasilocal} V. Faraoni, {\em Classical Quantum Grav.} {\bf 
33}, 015007 (2015); V. Faraoni and J. C\^ot\'e, {\em Phys. Rev. D} {\bf 
100}, 084015 (2019).

\bibitem{useful1} J.C. Hwang, {\em Phys. Rev. D} {\bf 42}, 2601 (1990); 
J.C. Hwang, {\em Class. Quantum Grav.} {\bf 7}, 1613 (1990); 
J.C. Hwang, {\em Class. Quantum Grav.} {\bf 8}, 195 (1991); 
J.C. Hwang, {\em Class. Quantum Grav.} {\bf 14}, 3327 (1997);
J.C. Hwang H. and Noh, {\em Phys. Rev. D} {\bf 54}, 1460 (1996).

\bibitem{useful2} M. Salgado, {\em Class. Quantum Grav.} {\bf 23}, 4719 
(2006).

\bibitem{Wald} R.M. Wald, {\em General Relativity} (Chicago 
University Press, Chicago, 1984).

\bibitem{reviews} T.P. Sotiriou and V. Faraoni, 
{\em Rev. Mod. Phys.} {\bf 82},  451 (2010); 
A. De Felice and S. Tsujikawa, 
{\em Living Rev. Relat.} {\bf 13}, 3 (2010);
S. Nojiri and S.D. Odintsov, 
{\em Phys. Rept.} {\bf 505}, 59  (2011). 

\bibitem{CMB} T. Clifton, D.F. Mota, and J.D. Barrow, {\em Mont. Not. Roy. 
Astron. Soc.} {\bf 358}, 601 (2005).

\bibitem{ourCMB}V. Faraoni, V. Vitagliano, T.P. Sotiriou, and S. Liberati, 
{\em Phys. Rev. D} {\bf 86}, 064040 (2012).

\end{thebibliography}
\end{document}